\documentclass[letterpaper, 10 pt, conference]{ieeeconf}
\usepackage{graphicx}
\usepackage{amsmath}
\usepackage{authblk}
\usepackage{hyperref}
\usepackage{dblfloatfix}
\usepackage{lipsum}
\usepackage{fancyhdr}
\usepackage{lastpage}
\usepackage{longtable, supertabular}

\IEEEoverridecommandlockouts                              
\title{AttoSats: ChipSats, other gram-scale spacecraft, and beyond}

\begin{document}

\renewcommand\Authfont{\small}
\renewcommand\Affilfont{\itshape\footnotesize}
\vspace{-10mm}


\twocolumn[{
\begin{center} 
\vspace{5mm}
\textbf{AttoSats: ChipSats, other Gram-Scale Spacecraft, and Beyond}\\ 
\vspace{5mm}
\text{Andreas M. Hein$^{a}$*}, \text{Zachary Burkhardt $^{a}$}, \text{Marshall Eubanks $^{a}$}\\  
\end{center}
$^{a}$ \emph{Initiative for Interstellar Studies}\\ \hspace{1.5mm}  [andreas.hein@i4is.org]\\ 
 * Corresponding Author
\begin{center}
\section{}\textbf{Abstract}
\end{center} 
The miniaturization of electronic and mechanical components has allowed for an unprecedented downscaling of spacecraft size and mass. Today, spacecraft with a mass between 1 to 10 grams,  AttoSats, have been developed and operated in space. Due to their small size, they introduce a new paradigm in spacecraft design, relying on agile development, rapid iterations, and massive redundancy. However, no systematic survey of the potential advantages and unique mission concepts based on AttoSats exists. This paper explores the potential of AttoSats for future space missions. First, we present the state of the art of AttoSats. Next, we identify unique AttoSat characteristics and map them to future mission capabilities. Finally, we go beyond AttoSats and explore how smart dust and nano-scale spacecraft could allow for even smaller spacecraft in the milligram range: zepto- and yocto spacecraft.

\hspace{0.5mm} \textbf{Keywords: ChipSat, small spacecraft, AttoSat, miniaturization, Starshot, Breakthrough Initiatives}
\vspace{5mm}
}]

\section{\textbf{1. Introduction}}

\begin{figure*}[htp]
    \centering
    \includegraphics[width=14cm]{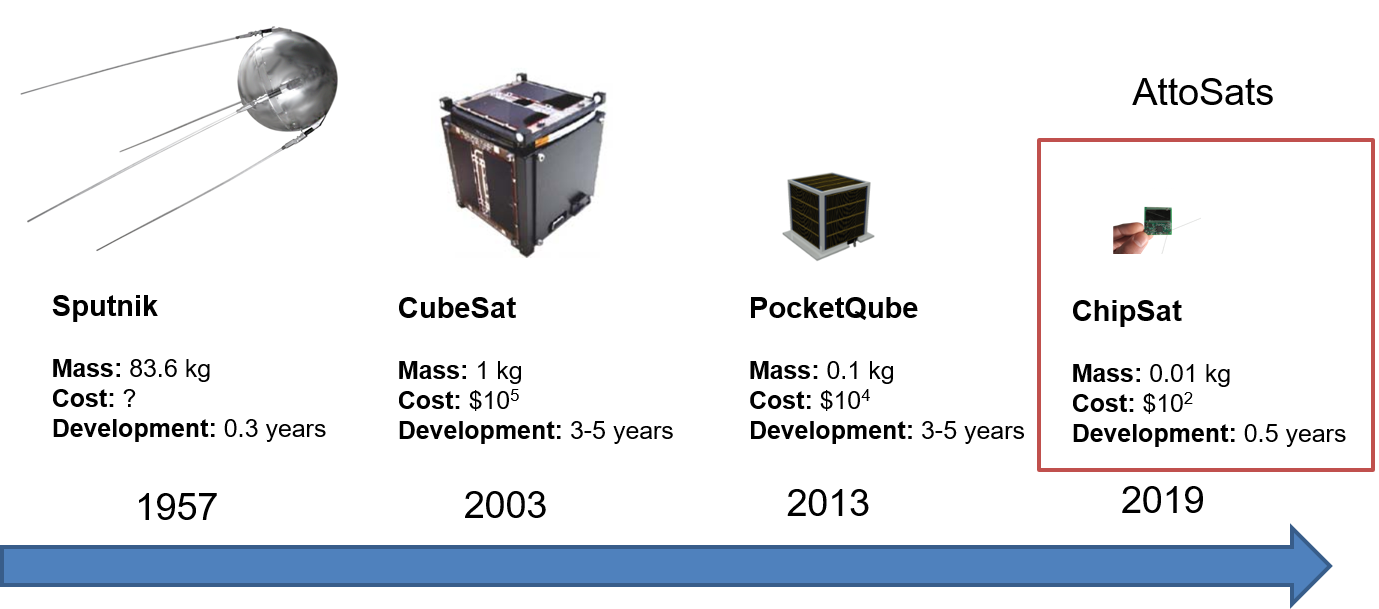}
    \caption{Evolution of spacecraft size}
    \label{fig:Size}
\end{figure*}

A prevailing trend of space systems is miniaturization with smaller and smaller spacecraft being developed, as shown in Figure \ref{fig:Size}. Most prominently, CubeSats (kg-scale spacecraft) have recently made spacecraft development accessible to universities and start-ups due to their low cost and quick development duration compared to traditional spacecraft. However, the development of CubeSats still costs on the order of $\$10^5 - \$10^6$ and takes several years to get from concept development to launch. Developing a CubeSat at a university still requires a considerable effort in terms of fundraising and project management, as students are rarely staying with the development team over its entire life cycle. The introduction of PocketQubes (5x5x5 cm unit(s)) and FemtoSats (100 gram-class spacecraft) has not fundamentally changed the effort required for developing the spacecraft \cite{Janson, Perez2016, Tahri2013}.\\

\begin{figure}[htp]
    \centering
    \includegraphics[width=8cm]{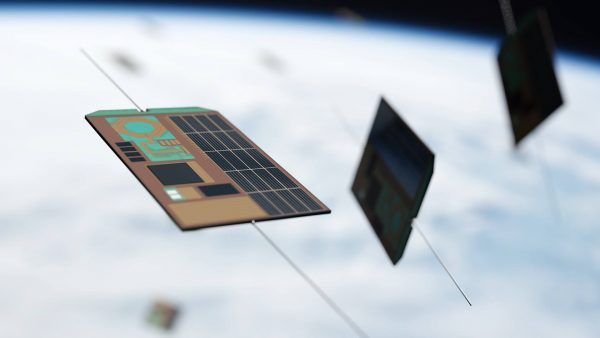}
    \caption{Artist's rendition of an AttoSat (Credit: Efflam Mercier)}
    \label{fig:Chip}
\end{figure}

More recently, a much smaller class of spacecraft has been introduced: ChipSats \cite{Manchester2015,Barnhart2009,Barnhart2007,Barnhart2006a,Barnhart2008a,Janson1999}. ChipSats are microchip-shaped spacecraft where all spacecraft components are integrated on a single PCB board or microchip, as shown in Figure \ref{fig:Chip}. The mass of these spacecraft is on the order of a few grams to 10s of grams. Hence, at the lower end of the mass range, they belong to the class of AttoSats (1 - 10 gram-class spacecraft). 

\begin{figure*}[htp]
    \centering
    \includegraphics[width=14cm]{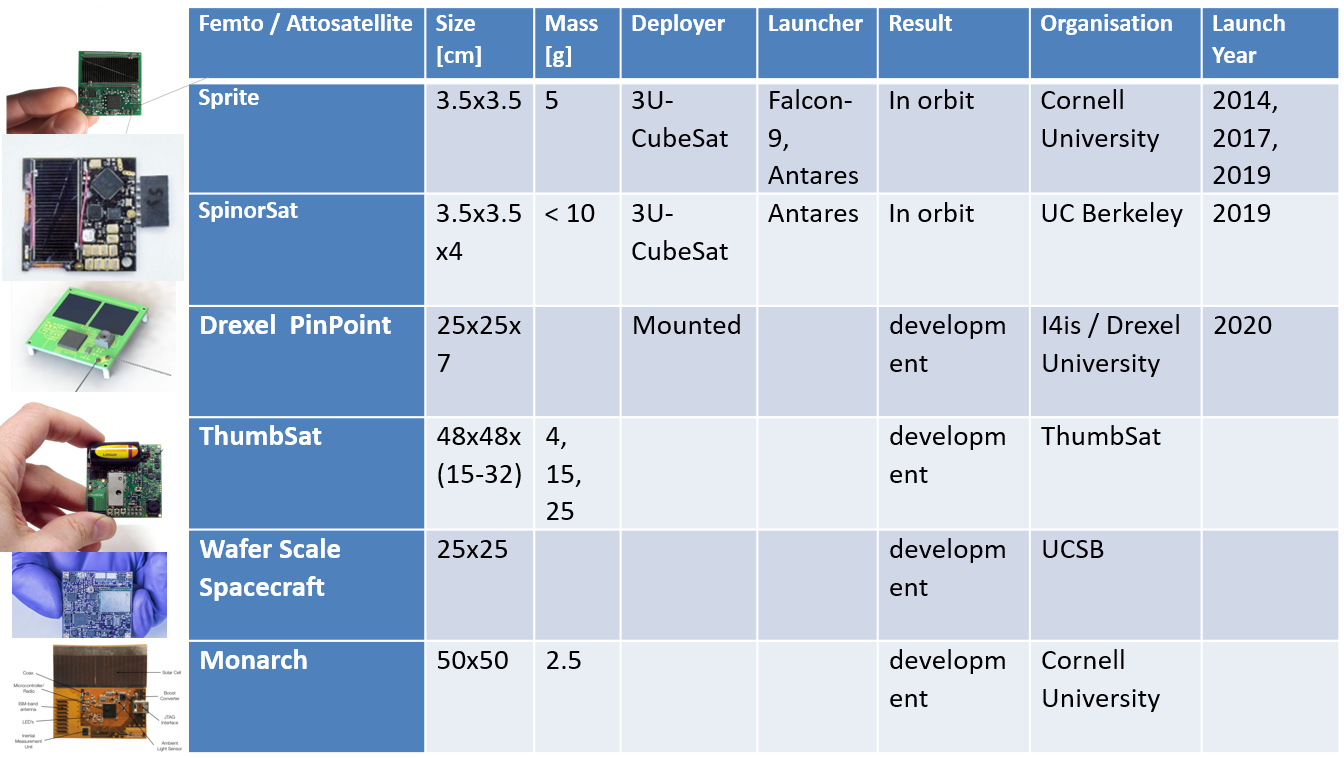}
    \caption{ChipSats launched and under development (Image credit: Cornell University, UC Berkeley, Drexel University, ThumbSat, UC Santa Barbara)}
    \label{fig:projects}
\end{figure*}

ChipSats and AttoSats have recently come to attention due to the KickSat project, where funding was raised via a Kickstarter campaign and about a hundred ChipSats could be aquired for the price of a mobile phone \cite{Manchester2013,Manchester2015}.  The recent lauch of KickSat-2 with the deployment and operation of ChipSats has demonstrated that ChipSats can be operated in space and send data back \cite{Krebs2016}. Furthermore, the Breakthrough Starshot project, announced in 2016, proposed the launch of a gram-scale spacecraft to another star within the next decades \cite{Lubin2016,Parkin2018}. \\

AttoSats are worth investigating, as they introduce a paradigm change in spacecraft development. The drastically reduced size and mass (2-3 orders of magnitude below CubeSats) reduces their development cost and effort drastically. ChipSat hardware costs on the order of dozens to hundreds of dollars and their development from concept to flight hardware can be finished within 6 months (Source: Zachary Manchester - conversation during FemtoSat workshop). This drastic reduction has several consequences: Spacecraft development now becomes accessible to almost anybody with basic knowledge in building electronic devices, the cost is on the order or less than for a mobile phone and therefore accessible to almost anybody who can afford a mobile phone. Furthermore, the drastically lower cost and mass allows for massive redundancy \cite{Manchester2011,Adams2019}. The short development duration has the potential for fast build-and-launch cycles, where new designs can be quickly iterated and directly tested in space. \\

In the following, we use the notion of \textit{AttoSat} whenever we refer to a spacecraft with a mass between 1 and 10 grams, independently of its shape. We use the notion of \textit{ChipSat} whenever the AttoSat has, in addition to its low mass, a flat shape and consists of a single printed circuit board (PCB). This distinction is similar to the one between PicoSats and CubeSats, where the former indicates spacecraft in a specific mass range and the latter refers to cube-shaped spacecraft. \\ 

\begin{figure*}[htp]
    \centering
    \includegraphics[width=16cm]{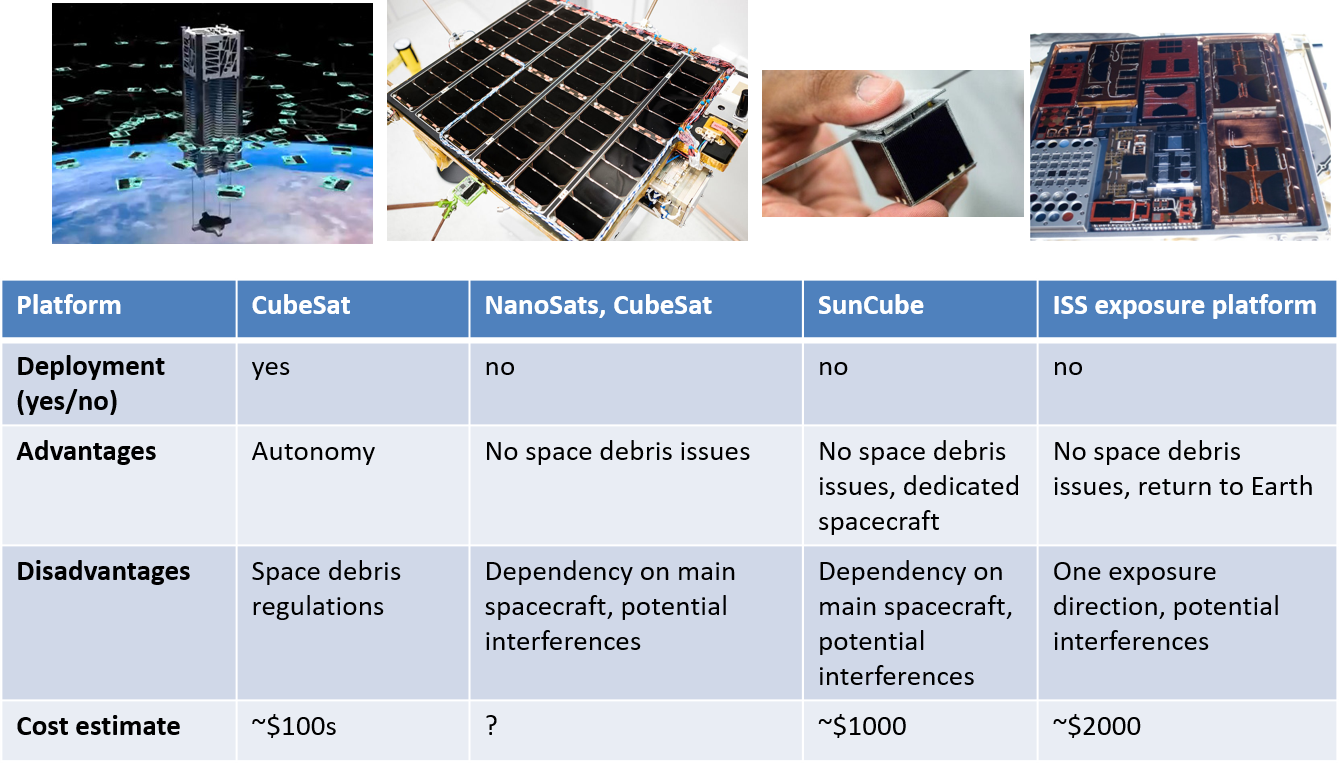}
    \caption{Overview of alternatives for ChipSat space exposure (Image credit: Cornell University, OHB, Arizona State University)}
    \label{fig:deployment}
\end{figure*}

Previous publications on AttoSats have mostly focused on their hardware design \cite{Manchester2015}, communication link design \cite{Manchester2015}, feasibility \cite{Barnhart2006a}, propulsion \cite{Atchison2010,Atchison2009,Atchison2007,Lubin2016,Streetman2009,Weis2014,Atchison2008,Burkhardt2018}, use for planetary exploration \cite{Streetman2016,Streetman2015}, orbit dynamics \cite{Colombo2013,Weis2014,Streetman2017,Weis2016,Weis2016a,Lucking2012,Atchison2010,Atchison2011}, atmospheric entry \cite{Atchison2010a,Vivenzio2019}, deployable structures for solar and laser sails \cite{Kaluthantrige2019,Rocha2019}, and impact of redundancy on mission design \cite{Adams2019,Manchester2011}. However, one of the open questions is, which missions would actually be enabled or drastically improved via ChipSat / AttoSat capabilities. In other words, their value proposition. \\

This paper presents a systematic assessment of ChipSat and AttoSat capabilities in order to identify promising missions for which their capabilities could make a substantial and unique contribution.

\section{\textbf{2. State of the Art}}
Several AttoSat development projects are currently under way with different deployment strategies. In the following, an overview of the state of the art of ongoing hardware development projects is provided. \\

An overview of launched ChipSats and ChipSats under development is provided in Figure \ref{fig:projects}. It can be seen that two ChipSat types were launched into space, the Sprite and SpinorSat. Both were deployed from a 3U-CubeSat. Further ChipSats are currently under development, such as the Drexel PinPoint, which is going to be mounted on a small satellite and not deployed. For the ThumbSat \cite{ThumbSat2015}, wafer-scale spacecraft \cite{Brashears2016}, and Monarch \cite{Adams2019}, the launch date and deployment mode is not known.\\

In contrast to larger spacecraft, not all ChipSats have been deployed in space. Some have been mounted on larger spacecraft. An overview of how ChipSats have been launched into space is provided in Figure \ref{fig:deployment}. A proven form of deployment is the release from a CubeSat. Releasing a large number of ChipSats from a CubeSat may be subject to additional regulatory scrutiny, which may lead to delays. As an alternative, ChipSats may be attached to a larger spacecraft but not released in space and stay attached. Larger spacecraft could be small spacecraft, CubeSats, or even femto-satellites such as the recently proposed SunCube \cite{Thangavelautham2016}. This has been done for Sprite ChipSats on the OHB Max Valier spacecraft. An advantage of this approach is that space debris mitigation issues are avoided. However, as the ChipSat is attached to the main spacecraft, dependencies are introduced, such as in terms of orientation and communication frequencies. Last but not least, Sprite ChipSats have first been tested in space on an exposure platform on the International Space Station (ISS). Exposure platforms have the advantage that space debris issues are not present and the ChipSats can be returned to Earth for inspection. On the downside, exposure platforms on the ISS face a single direction, which might introduce limitations in terms of ChipSat operations, in particular regarding solar radiation influx, in case the platform faces the Earth. Also, electromagnetic interference with other equipment on the ISS needs to be avoided. 
\section{\textbf{3. Attosat Characteristics and Capabilities}}
We introduce the characteristics - capability method for a structured approach to connect unique AttoSat characteristics with capabilities which are enabled by them. \\

\subsection{Characteristics - capabilities method}
\begin{figure}[htp]
    \centering
    \includegraphics[width=8cm]{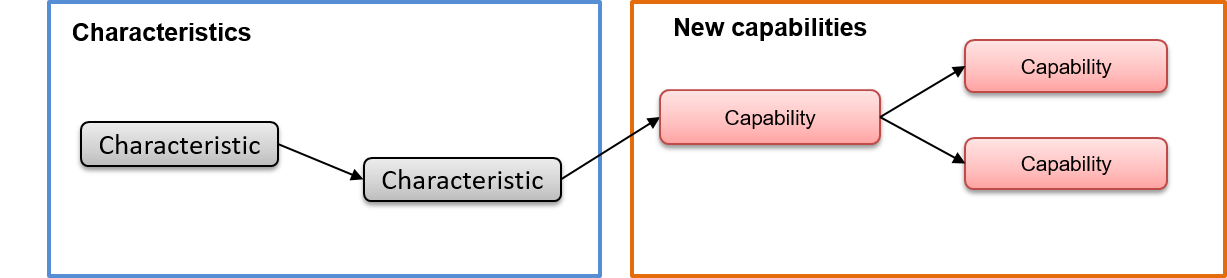}
    \caption{Characteristics and capabilities}
    \label{fig:char_basic}
\end{figure}

The characteristics - capabilities method first asks "What are specific characteristics of the technology?". These are characteristics that are considered "unique" or "distinguishing" compared to other technologies. For example, AttoSats have, by definition a lower mass than PicoSats, as shown in Figure \ref{fig:char_basic}. There are characteristics that may follow from this characteristic. Lower mass means that more satellites can be launched into space for the same available payload mass. Some of these characteristics might be exploited for doing something new or better than before. This is where capabilities come in.\\

\begin{figure}[htp]
    \centering
    \includegraphics[width=8cm]{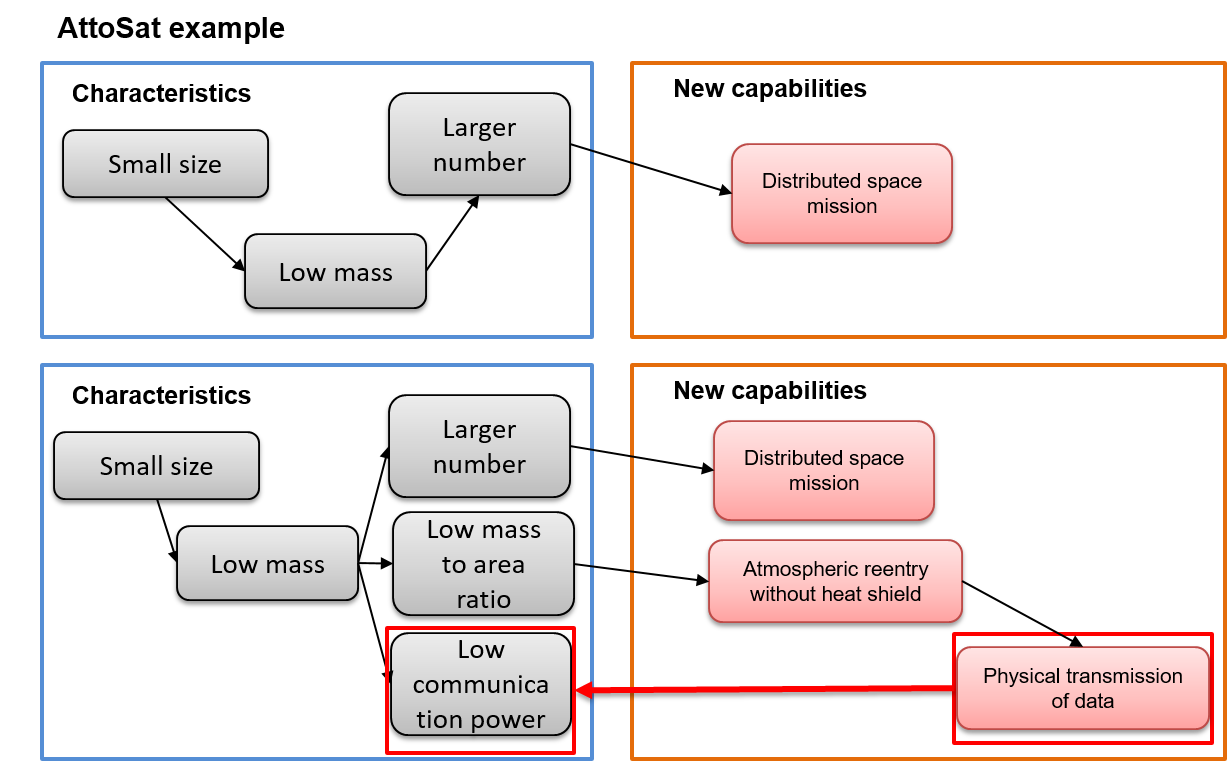}
    \caption{AttoSat sample characteristics and capabilities}
    \label{fig:char_atto}
\end{figure}

The potential of doing something is called \textit{capability} in the following \cite{Hein2016c}. In the second step of the characteristics - capabilities method, capabilities are derived from characteristics by asking the question "What can be done in a new way or better way by exploiting one or more characteristics?". Launching a \textit{larger number of satellites} (characteristic) into space enables \textit{distributed space missions} (capability), as shown in the upper part of Figure \ref{fig:char_atto}. Creativity methods such as TRIZ \cite{Savransky2000} can be used for generating ideas for capabilities. \\
Furthermore, some characteristics of the technology might be considered undesirable, i.e. negative. The small size of AttoSats implies that the power generated on board is very limited, which means that available power for communications is also very limited. A question which can be asked in such a case is "How can I exploit characteristics and capabilities for countering disadvantages?" A possibility could be to exploit the lower mass to area ratio to allow for atmospheric entry without a heat shield and physically transfer data via de-orbiting the AttoSat (Idea proposed by Mason Peck). This is shown in the lower part of Figure \ref{fig:char_atto}. The red arrow indicates that the capability "Physical transmission of data" may remedy for the disadvantageous characteristic "Low communication power". Of course, the remedy might not only be a remedy but actually become a major advantage for the technology. Physical storage of data might be on the order of gigabytes and thereby be much higher then via remote communication. Compensating for negative characteristics can be done systematically by identifying technical or physical contradictions from TRIZ and solving these using one of the solution principles such as separation in space and time \cite{Savransky2000}. However, other creativity methods may also be used. \\

In the following, we collected characteristics and capabilities from the existing literature or developed them during discussions with members of the Initiative for Interstellar Studies and people external to the organization.    

\subsection{AttoSat characteristics}
\begin{figure*}[htp]
    \centering
    \includegraphics[width=18cm]{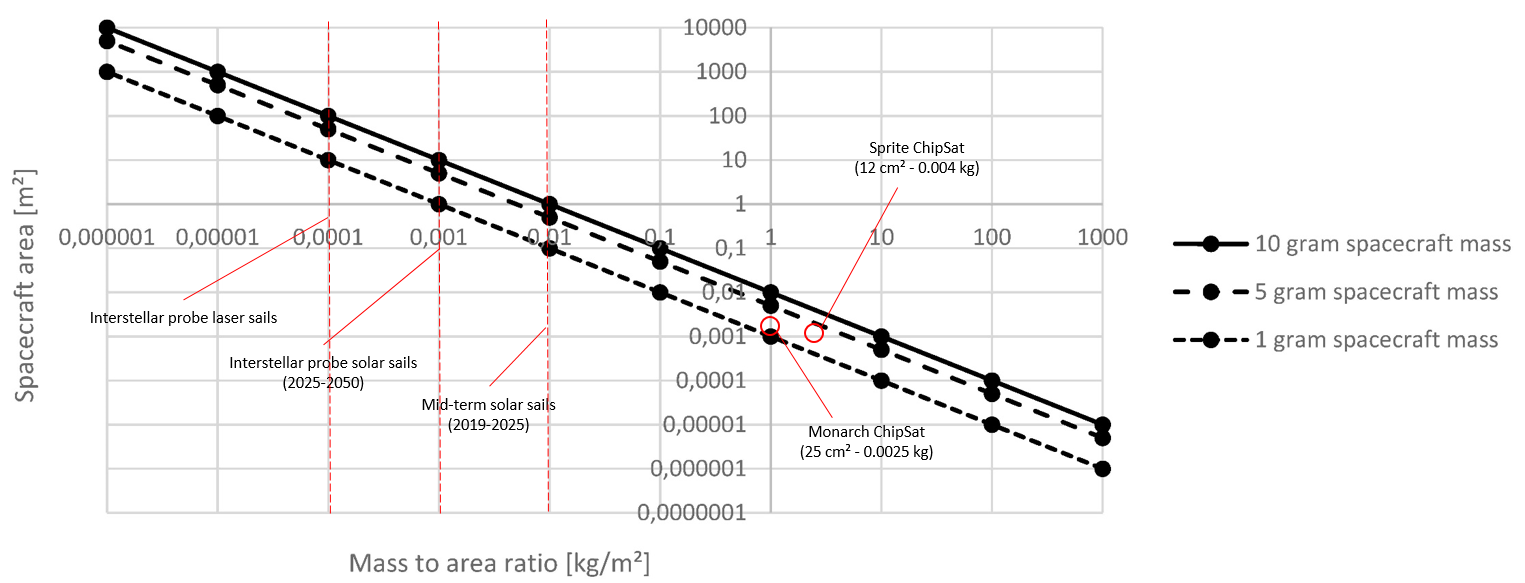}
    \caption{AttoSat surface area with respect to mass to area ratio and mass}
    \label{fig:mta}
\end{figure*}

\begin{figure*}[htp]
    \centering
    \includegraphics[width=18cm]{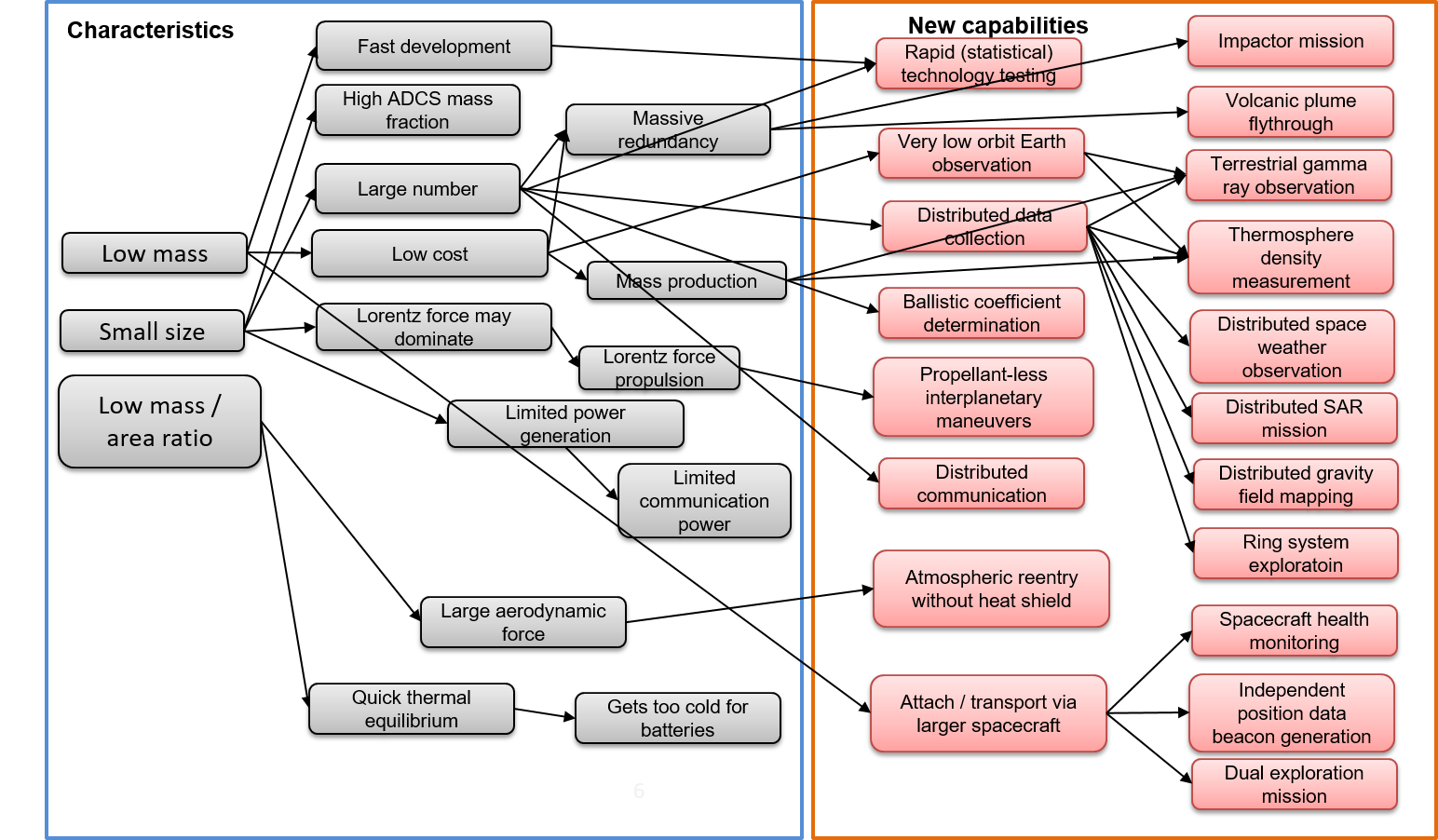}
    \caption{AttoSat characteristics and capabilities}
    \label{fig:Char}
\end{figure*}

In the following, we will apply the characteristics - capabilities method to the case of AttoSats and start with AttoSat characteristics. The main distinguishing, physical characteristics of AttoSats directly follow from their definition and are shown on the far left of Figure \ref{fig:Char}. First of all, it is their low mass (1-10 grams). Second, AttoSats usually have a small size, assuming that their density is relatively high, i.e. they are tightly integrated. The small size may translate into a low mass to area ratio, in case the AttoSat is flat, such as in the case of a ChipSat. Figure \ref{fig:mta} shows ranges of AttoSat surface areas with respect to different mass to area ratios and AttoSat masses. Sample values for the Sprite ChipSat \cite{Manchester2015} and Monarch ChipSat \cite{Adams2019} are given. Furthermore, values for solar and laser sail AttoSats are given on the right side. The mass to area values correspond to those given in Parkin \cite{Parkin2018} and Montgomery and Johnson \cite{Montgomery2017}. \\

From these characteristics follow further characteristics. The low mass correlates with fast development, due to the low number of components and low complexity \cite{Manchester2015}. Low mass also correlates with low cost in the space domain \cite{Larson1999a}. The small size correlates with a high mass fraction for the attitude determination and control system (ADCS) \cite{Streetman2017}. The small size, however, allows for packing a larger number of spacecraft into a given payload volume, leading to a potentially large number of spacecraft launched into space. From this follows that massive redundancy can be established \cite{Adams2019,Manchester2011}. Furthermore, AttoSats can be mass-produced, thereby exploiting economy-of-scale effects. \\

The small size also means that certain physical effects are larger. This is the case for the Lorentz force \cite{Atchison2011}. The larger Lorentz force compared to perturbations such as aerodynamic force can be exploited in the form of Lorentz force propulsion \cite{Atchison2011,Atchison2009,Atchison2007,Streetman2009}. The small size also means that the surface area is limited, although lower mass to area values may remedy for this. In other words, the spacecraft would be flatter and have a larger surface. However, otherwise, the limited surface area means that only a fraction of the power available for PicoSats can be generated on an AttoSat. This obviously constrains subsystem power supply such as for communications, which is a key limiting factor for very small spacecraft. On the other side, a low mass to area ratio means that aerodynamic forces are larger relative to spacecraft mass than for PicoSats. However, the low mass to area ratio also means that a large surface area is either facing the Sun or open space and therefore, the thermal variations are potentially large \cite{Bruno2014}. In particular, the spacecraft tends to cool down significantly, making the use of existing batteries infeasible \cite{Manchester2015}.  

\subsection{AttoSat-enabled capabilities}
The AttoSat characteristics from the previous section are used for identifying new capabilities, i.e. things that can be done in a new or better way. Several capabilities have already been reported or developed in the literature. However, we also present capabilities which, to our knowledge, have not yet been reported. \\
The characteristic of a large number of AttoSats translates into several capabilities. First, spatially and temperally distributed data collection becomes possible. A potential mission capability would be terrestrial gamma ray observation. Terrestrial gamma ray flashes (TGFs) occur in Earth's atmosphere at altitudes between 10 and 20 km during lightning strikes. They generate high energy electrons and positrons that travel upwards into space. These particles are detectible as high as 500 km. AttoSat compatible CdZnTe detectors could distinguish TGF photons from background noise with 2 orders of magnitude margin. Such an AttoSat swarm could greatly increase knowledge of TGFs, their particularly frequency of occurrence, geographic distribution, and energy range. As AttoSats are low cost, very low orbits, where spacecraft may only operate for days to weeks and still remain cost-effective. \\

Another potential mission capability is the measurement of the thermosphere density. The density of the lower thermosphere (80 to 350 km) is badly characterized, as it highly dependent on solar radiation. Again, distributed measurements at very low orbits are feasible, as the AttoSats are low cost and allow for distributed measurements. \\

Space weather observations also exploit the capability of distributed data collection, such as characterizing plasma bubbles. These are areas of depleted density in ionospheric plasma, which can cause deflection of communications signals. Another space weather phenomenon is the magnetotail, which is the extension of Earth’s magnetic field on the side facing away from the sun. the magnetotail is heavily influenced by incident solar wind.\\ 

Distributed synthetic aperture radar (SAR) applications are also possible using distributed spacecraft. Gravity field mappings could also be conducted in a more efficient way by distributed spacecraft, e.g. at the vicinity of small Solar System bodies. \\

Furthermore, exploring planetary or small-body ring systems could also be conducted efficiently via distributed AttoSats. \\

The massive redundancy could be exploited in impactor missions such as on small bodies or moons \cite{Adams2019,Streetman2016,Manchester2011} or flying through volcanic plumes of moons such as Europa and Enceladus. In the former case, redundancy is exploited in terms of survivability. Even in case most of the AttoSats do not survive the impact, a sufficient number may survive to transmit data back from the surface. In the latter case, redundancy is exploited to increase the probability that at least one AttoSat is subject to a rare event, in this case flying through the eruption of a cryovolcano. \\

The characteristic of AttoSats in large numbers allows for the capability of measuring a wide range of ballistic coefficients for spacecraft. \\

Lorentz force propulsion allows for unprecedented maneuvering capabilities of AttoSats for interplanetary travel \cite{Atchison2009,Atchison2007,Streetman2009}. \\

Due to the potentially low mass to area ratio of AttoSats, Atchison et al. \cite{Atchison2010a} and Vivenzio et al. \cite{Vivenzio2019} proposed atmospheric entry without a heat shield. This capability can be exploited for atmospheric measurements or deploying light-weight sensor networks on planets or moons with atmospheres.\\
The characteristic that AttoSats can be developed quickly, even within a few months, allows for rapid technology testing in space, primarily of small components. Technologies could be matured and iterated quickly. A precondition is that the spacecraft can be launched in quick succession. The large number of spacecraft would potentially add the dimension of statistical testing by getting multiple data points for the same type of component but tested on several spacecraft. \\
The low mass also allows for attaching AttoSats to and transporting on a larger spacecraft. This capability could be used for spacecraft health monitoring by observing the main spacecraft. It could also be used for generating position data while being attached to the larger spacecraft. Finally, dual exploration missions may become possible, where the main spacecraft deploys a large number of AttoSats for additional data collection \cite{Streetman2016,Streetman2015}.  \\
In the following, we will describe in more detail some of the missions, based on these capabilities. 

\subsection{Sample AttoSat-enabled missions}
\subsubsection{Distributed Communications}
AttoSats could enable a new type of distributed space communications mission. The low cost and mass producibility of AttoSats could be leveraged to create a constellation consisting of a very large number of satellites. This constellation would receive communications signals from one ground station and forward it through the network of AttoSats to another location. \\
This approach offers a number of possible advantages over other approaches to space-based telecommunications. Unlike other low orbiting communications constellation concepts, an AttoSat constellation could use fixed ground station antennas; with a sufficient number of AttoSats in the constellation there would consistently be a spacecraft in the antenna's field of view. This would significantly reduce the tracking and ground station requirements for the constellation. \\
This concept could be deployed as a standing constellation or enable rapid establishment of a temporary AttoSat constellation. A future small launch system optimized for low mass vehicles of AttoSats could quickly deploy a dedicated communications constellation of AttoSats to connect a small set of particular interest areas. Such a constellation could play an important role in restoring communications networks in response to a disaster or failure of more traditional communications systems. \\

\subsubsection{AttoSat Synthetic Aperture Radar}
A small swarm of AttoSats could enable a new approach towards synthetic aperture radar (SAR) missions. Such a mission would use multiple radar receiver AttoSats to achieve a bi-static SAR configuration. A bi-static SAR system uses one emitter, in this case carried by the CubeSat with multiple receivers enabling reflected radar waves to be detected from multiple locations at once. This architecture could potentially enable equivalent or better SAR performance at a fraction of the cost. Wang et al. \cite{Wang2017} estimate that a swarm with 50 mico- and nano-satellites could achieve a high enough resolution to identify ships and airplanes.\\

\subsubsection{Spacecraft Health Monitoring}
Many recent studies have examined the possibility of using secondary spacecraft for inspection or servicing of large spacecraft. AttoSats represent an ideal candidate for the inspection portion of such concepts. Previous technology demonstration efforts for satellite inspection concepts have identified the potential for collision with the host satellite as a significant technical risk. This risk is increased by the need to keep the inspection satellite within the vicinity of the host spacecraft for long periods to facilitate continued inspection. Using AttoSats would mitigate this risk. The low cost and mass of AttoSats would allow multiple inspection spacecraft to be co-manifest with a large mission; these AttoSats could then be deployed onto trajectories that remove them from the vicinity of the host satellite relatively quickly. The inspectors could be as needed over the lifetime of the mission, providing necessary inspections without the need for a cooperating spacecraft. Because of the properties of AttoSats, these inspectors could potentially be standardized and developed at a very low cost enabling them to be included with a wide variety of missions. 

\subsubsection{Very-Low Orbit Science Missions}
Many open science questions could be studied by spacecraft placed into very low Earth orbits. In the past, these questions have been hard to address because the lifetime of spacecraft in such orbits is usually very short. AttoSats present a potential solution for implementing missions that can take meaningful measurements in Earth's lower thermosphere. Because of their low cost, the short mission lifetime in low orbits is less of a concern for an AttoSat mission. Additionally, the ability to deploy a large swarm of AttoSats would provide the potential to take simultaneous measurements at a variety of locations around the Earth. Alternatively, AttoSats could be used to take measurements at desired times by deploying the swarm into a higher orbit and having AttoSats drop into lower orbits when desired using small scale propulsion systems or by varying the attitude of the spacecraft to control drag. Scientific investigations that could be accomplished with a low orbiting AttoSat mission include studies of the density of the lower thermosphere, terrestrial gamma ray flashes, plasma bubbles, and space weather phenomena in Earth's magnetotail. \\

\subsubsection{Rapid (statistical) technology testing}
The rapid increase in CubeSats under development has also lead to an increased demand for testing components in a space environment. However, testing small components, for example, a new Sun sensor, in space is challenging, due to the lack of flight opportunities and the high cost of developing a dedicated CubeSat for this purpose. AttoSats may provide a platform for testing small components in space. Due to their small size, they can be developed at a low cost and due to their low mass, more launch opportunities might be available. The component to be tested could be quickly iterated and tested in space. Furthermore, multiple AttoSats with the same component could be launched into space in order to get rapid, statistical data for multiple components. This application for AttoSats is currently under investigation at the Technical University of Munich (Source: Martin Langer and Martin Dziura). \\

\subsubsection{Ballistic coefficient determination}
Space debris have become a major issue for low orbits around Earth. Existing simulations for the dynamics of space debris rely on approximations for atmospheric densities and the aerodynamic characteristics of space debris. A key aerodynamic parameter is the ballistic coefficient, as it determines how fast space debris will deorbit. AttoSats could be used for determining the ballistic coefficient of space debris under various atmospheric densities and thereby lead to more precise space debris simulations. Various shapes would be attached to the AttoSats, simulating different space debris shapes. The orbital decay data would then be used to infer the orbital decay behavior of space debris. Measurements at different points in time at various atmospheric densities would add to the precision of future space debris simulations.\\   

\subsubsection{Position data beacon}
Today, spacecraft and space debris are tracked via radar. However, the precision of the position data could be significantly improved by an additional beacon on board of the spacecraft. An AttoSat which would be attached to a larger spacecraft could act as a beacon. The AttoSat would not be structurally independent, as it is attached to the main spacecraft. However, it would be otherwise an independent spacecraft, e.g. with its own power supply and communications subsystem.

\section{\textbf{3. Future Prospects}}
\begin{figure*}[htp]
    \centering
    \includegraphics[width=14cm]{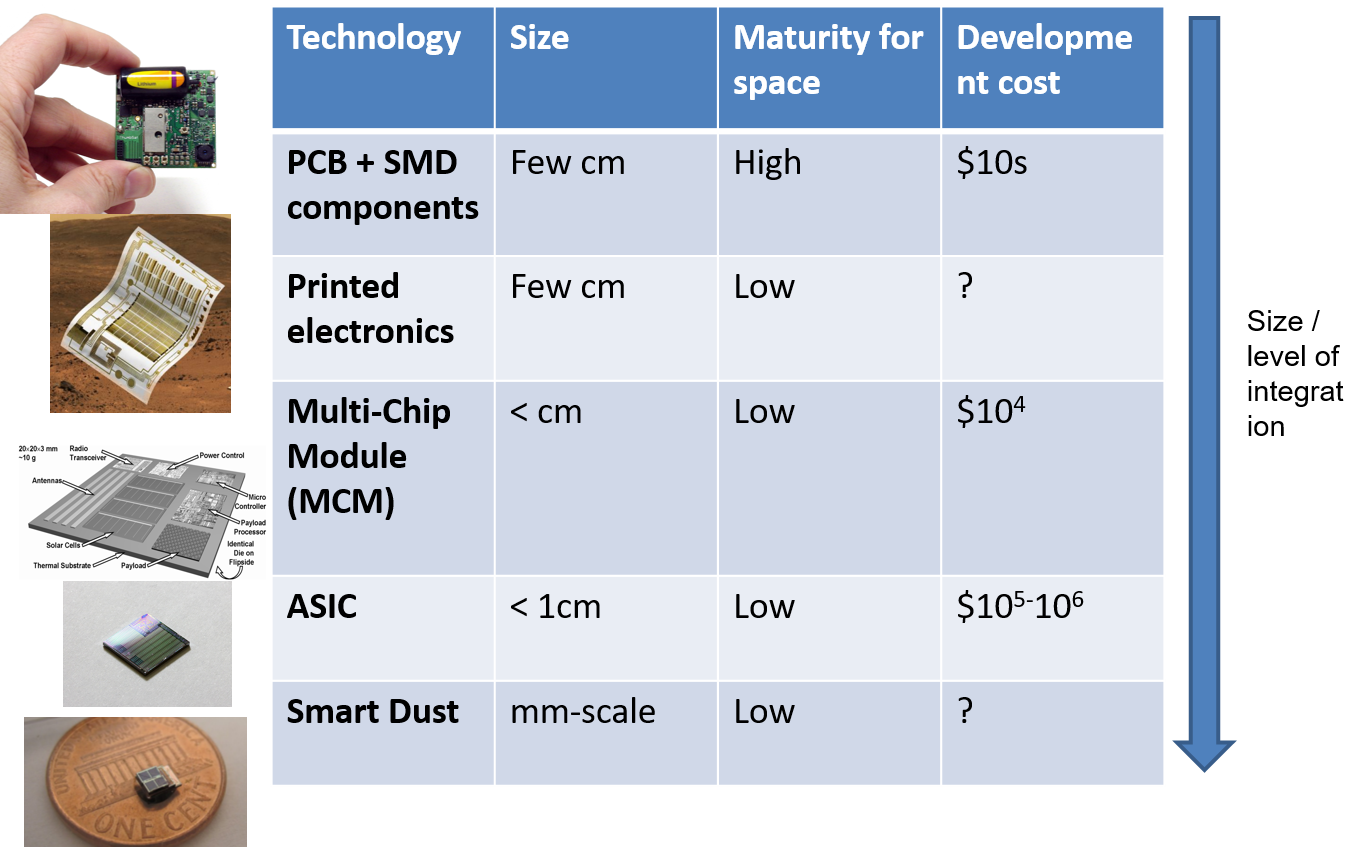}
    \caption{AttoSat technologies for further miniaturization}
    \label{fig:technology}
\end{figure*}

Due to the small number of existing AttoSat projects, it is difficult to identify trends. There are two parameter values that may evolve independently: size and mass. The upper limit to mass is 10 grams. With the launched Sprite ChipSats having a mass of about 4 grams \cite{Manchester2015}, there is the potential of heavier AttoSats. \\
Regarding the size of AttoSats, a possible trend is towards lighter but flatter AttoSats. The Monarch ChipSat prototype, currently under development at Cornell University, has a mass of 2.5 grams \cite{Adams2019} but uses a significantly lighter substrate than the Sprites. Size-wise ChipSats with roughly the same mass as the KickSat ChipSats are currently under development at Cornell University. The Monarch ChipSat has a surface area of 25 $cm^2$ and are significantly larger than the Sprite Chipsat with an area of 12.25 $cm^2$. As shown in Figure \ref{fig:mta}, AttoSats may further evolve into the direction of lower mass to area ratios, in particular for solar and laser sail applications. Parkin \cite{Parkin2018} proposes point designs for laser sail interstellar missions with a mass to area ratio between 0.2 $g/m^2$ and 0.55 $g/m^2$.\\
Another possible direction in AttoSat evolution could be towards higher integration into a given volume. This would be just an extrapolation of the current satellite miniaturization trend where smaller satellites tend to have higher densities. For example, the Initiative for Interstellar Studies has proposed the \textit{Midge} AttoSat, based on a $\mu$duino board which would be vertically integrated. Going beyond existing PCB technology, current trends in three dimensional integrated circuits may allow for vertically stacked AttoSats \cite{Topol2006}. One major drawback is the correspondingly smaller surface area. Therefore, power generation via solar cells is going to be difficult, unless they are deployed. \\
The potential of further miniaturization is constrained by the underlying electronics technology and factors related to spacecraft engineering. Regarding the underlying electronics technology, current AttoSats are primarily based on PCB technology with surface mounted devices (SMD), which are soldered to the PCB. This technology is limited by the size of SMD components and the circuit density on the PCB. \\
Several technologies exist that may overcome these limitations. An overview of these technologies is provided in Figure \ref{fig:technology}. Printed electronics may enable a further decrease in the thickness of the AttoSat. This technology was explored for a distributed sensor network for planetary exploration \cite{Short2014}. \\
As already proposed by Barnhart \cite{Barnhart2006}, system-on-a-chip technology could be used for further miniaturization. A system-on-a-chip integrates all spacecraft subsystems onto a single silicon substrate.

In contrast to PCB technology, the spacecraft no longer consists of individual integrated circuits (ICs) but comprises a single IC \cite{Janson1999}. Furthermore, the IC is purpose-built and its development is several orders of magnitude more costly than for PCBs, mainly due to their tight integration and customization. Possible technologies are multi-chip modules (MCMs) and application-specific integrated circuits (ASICs). MCMs are multiple ICs which are integrated in a single substrate. The development cost is on the order of $\$10^4$ and their size can be on the order of 1 cm. By contrast, ASICs are purpose-built and highly integrated. Radiation-hardened ASICs are typically used in automotive electronics. The StarChip proposed by Breakthrough Starshot seems to use ASIC technology. The development cost of ASICs is substantially higher than for MCMs and on the order of $\$10^5 - \$10^6$. \\ 

A more advanced technology is smart dust technology. Smart dust is a collection of mm-scale computing nodes with sensors \cite{Pister1999,Doherty2001,Warneke2001a,Warneke2001}. Power is either provided externally, e.g. via a laser or by solar cells and a battery \cite{Doherty2001}. The use of smart dust for space applications has recently been explored by Niccolai et al. \cite{Niccolai2019}. Smart dust has a mass below 1 gram and would therefore belong to the class of \textit{ZeptoSats}. \\
A key limitation of smart dust spacecraft is communications. Although radio and laser communication might have strong limitations due to power constraints, there are potential applications where data could be collected in smart dust devices and then downloaded to a larger spacecraft at close distance. One potential application would be the release of a smart dust cloud during the traversal of a scientifically interesting area. The collected data would be transmitted to the main spacecraft via e.g. passive laser communication, where a laser beam from the main spacecraft is reflected back from the smart dust nodes via a mirror. The mirror would opto-mecanically encode information into the reflected beam.\\
Regarding lower limits to spacecraft size, one can even go to much smaller scales. Several paths to further miniaturization of electronics are currently under investigation: 

\begin{itemize}
    \item Nanoelectronics \cite{Waser2003,Lu2010,Akinwande2014}
    \item Molecular scale electronics \cite{Xiang2016}
    \item Nanoelectromechanical systems (NEMS) \cite{Craighead2000}
    \item Micrometer-thin lenses \cite{Meem2019}
\end{itemize}

A more in-depth exploration of the potential of these technologies in spacecraft remains to be conducted. \\

Furthermore, recent advances in nanorobotics enabled micrometer scale robots. Currently, these robots are based on DNA molecules and operate in "wet" environments, e.g. within the human body \cite{Eelkema2006,Douglas2012}. Hence, their use in the space environment, at least in their current form is not feasible. An exception may be environments where liquids are present, such as on Europa and Enceladus.\\

Bottlenecks for further miniaturization of space systems is related to power and communications \cite{Hein2017}. Power for small spacecraft is mainly required for communications, as there are fundamental limits to how much power is needed for transferring one bit over a specific distance \cite{Messerschmitt2012}. A potential remedy might be to significantly decrease the mass to area ratio, at least for the antenna. Power might be generated by thin tethers, interacting with charged particles in interplanetary or interstellar space \cite{Matloff2005}. Although technically very challenging, there does not seem to be a principle obstacle for developing spacecraft with a mass below 1 gram or even in the milligram range. Such \textit{ZeptoSats} and \textit{YoctoSats} might open up yet unprecedented mission types.  

\section{\textbf{4.Conclusion}}
In this paper, we explored unique characteristics of AttoSats and how they may translate into unprecedented mission capabilities. We furthermore presented new mission concepts for AttoSats such as distributed communications, distributed synthetic aperture radar, spacecraft health monitoring, very low orbit science missions, rapid statistical technology testing, ballistic coefficient determination, and as position data beacons. Regarding future prospects, AttoSats with lower mass to area ratios may be promising, in particular for solar and laser sailing missions. Emerging electronics technologies may allow for a further miniaturization of spacecraft down to milligrams. Nevertheless, solving the power - communication bottleneck will be essential. For future work, we propose more in-depth studies of the presented mission concepts and how the power - communication bottleneck might be solved by exploiting the distributed nature of AttoSats and smaller spacecraft.

\graphicspath{ {images/} }

\vspace{0mm}



\section*{\textbf{Acknowledgments}}
We would like to thank Martin Langer and Barnaby Osborne for inspiring discussions on AttoSat use cases. Furthermore, we would like to thank Chris Welch from the International Space University and Pete Worden from Breakthrough Initiatives for involving i4is in the creation of the ChipSat study module.


\vspace{5mm}
\bibliographystyle{IEEEtran}
\bibliography{library}

\begin{thebibliography}{10}
\providecommand{\url}[1]{#1}
\csname url@samestyle\endcsname
\providecommand{\newblock}{\relax}
\providecommand{\bibinfo}[2]{#2}
\providecommand{\BIBentrySTDinterwordspacing}{\spaceskip=0pt\relax}
\providecommand{\BIBentryALTinterwordstretchfactor}{4}
\providecommand{\BIBentryALTinterwordspacing}{\spaceskip=\fontdimen2\font plus
\BIBentryALTinterwordstretchfactor\fontdimen3\font minus
  \fontdimen4\font\relax}
\providecommand{\BIBforeignlanguage}[2]{{%
\expandafter\ifx\csname l@#1\endcsname\relax
\typeout{** WARNING: IEEEtran.bst: No hyphenation pattern has been}%
\typeout{** loaded for the language `#1'. Using the pattern for}%
\typeout{** the default language instead.}%
\else
\language=\csname l@#1\endcsname
\fi
#2}}
\providecommand{\BIBdecl}{\relax}
\BIBdecl

\bibitem{Janson}
S.~W. Janson and D.~J. Barnhart, ``{The Next Little Thing : Femtosatellites},''
  in \emph{Small Satellite Conference 2013}, 2013.

\bibitem{Perez2016}
T.~R. Perez and K.~Subbarao, ``{A Survey of Current Femtosatellite Designs ,
  Technologies , and Mission Concepts},'' \emph{Journal of Small Satellites},
  vol.~5, no.~3, pp. 467--482, 2016.

\bibitem{Tahri2013}
N.~Tahri, C.~Hamrouni, and A.~M. Alimi, ``{Study of Current Femto-Satellite
  Approches and Services},'' \emph{International Journal of Advanced Computer
  Science and Applications}, vol.~4, no.~5, pp. 148--153, 2013.

\bibitem{Manchester2015}
Z.~Manchester, ``{Centimeter-Scale Spacecraft: Design, Fabrication, And
  Deployment},'' Ph.D. dissertation, 2015.

\bibitem{Barnhart2009}
D.~J. Barnhart, T.~Vladimirova, and M.~N. Sweeting, ``{Satellite
  Miniaturization Techniques for Space Sensor Networks},'' \emph{JOURNAL OF
  SPACECRAFT AND ROCKETS}, vol.~46, no.~2, 2009.

\bibitem{Barnhart2007}
------, ``{Very-Small-Satellite Design for Distributed Space Missions},''
  vol.~44, no.~6, 2007.

\bibitem{Barnhart2006a}
D.~J. Barnhart, T.~Vladimirova, and A.~M. Baker, ``{A Low-Cost Femtosatellite
  to Enable Distributed Space Missions},'' vol. 298, no. 0704, 2006.

\bibitem{Barnhart2008a}
D.~J. Barnhart, ``{Very Small Satellite Design for Space Sensor Networks
  Submitted for the Degree of N / A},'' no. June, 2008.

\bibitem{Janson1999}
S.~Janson, ``{Mass-producible silicon spacecraft for 21st century missions},''
  in \emph{Space Technology Conference and Exposition}, 1999, p. 4458.

\bibitem{Manchester2013}
\BIBentryALTinterwordspacing
Z.~Manchester, M.~Peck, and A.~Filo, ``{Kicksat: A crowd-funded mission to
  demonstrate the world's smallest spacecraft},'' in \emph{27th Annual AIAA/USU
  Conference on Small Satellites}, 2013. [Online]. Available:
  \url{http://digitalcommons.usu.edu/smallsat/2013/all2013/111/}
\BIBentrySTDinterwordspacing

\bibitem{Krebs2016}
\BIBentryALTinterwordspacing
G.~Krebs, ``{KickSat 1, 2},'' 2016. [Online]. Available:
  \url{http://space.skyrocket.de/doc{\_}sdat/kicksat-1.htm}
\BIBentrySTDinterwordspacing

\bibitem{Lubin2016}
P.~Lubin, ``{A Roadmap to Interstellar Flight},'' \emph{Journal of the British
  Interplanetary Society}, vol.~69, no. 2-3, 2016.

\bibitem{Parkin2018}
K.~L. Parkin, ``{The Breakthrough Starshot System Model},'' \emph{Acta
  Astronautica}, vol. 152, pp. 370--384, 2018.

\bibitem{Manchester2011}
Z.~Manchester and M.~Peck, ``{Stochastic space exploration with microscale
  spacecraft},'' in \emph{AIAA Guidance, Navigation, and Control Conference},
  2011, p. 6648.

\bibitem{Adams2019}
\BIBentryALTinterwordspacing
V.~Adams and M.~Peck, ``{Chipsats: R-Selected Spacecraft},'' 2019. [Online].
  Available: \url{10.31224/osf.io/wxhpt}
\BIBentrySTDinterwordspacing

\bibitem{Atchison2010}
J.~Atchison and M.~Peck, ``{A passive, sun-pointing, millimeter-scale solar
  sail},'' \emph{Acta Astronautica}, vol.~67, no.~1, pp. 108--121, 2010.

\bibitem{Atchison2009}
J.~A. Atchison and M.~A. Peck, ``{Lorentz-Augmented Jovian Orbit Insertion},''
  \emph{Journal of Guidance, Control, and Dynamics}, vol.~32, no.~2, pp.
  418--423, mar 2009.

\bibitem{Atchison2007}
J.~Atchison and M.~Peck, ``{A millimeter-scale lorentz-propelled spacecraft},''
  in \emph{AIAA Guidance, Navigation and Control Conference and Exhibit}, 2007,
  p. 6847.

\bibitem{Streetman2009}
B.~Streetman and M.~Peck, ``{Gravity-assist maneuvers augmented by the Lorentz
  force},'' \emph{Journal of guidance, control, and dynamics}, vol.~32, no.~5,
  pp. 1639--1647, 2009.

\bibitem{Weis2014}
\BIBentryALTinterwordspacing
L.~Weis and M.~Peck, ``{Active Solar Sail Designs for Chip-Scale Spacecraft},''
  in \emph{AIAA/USU Small Satellite Conference 2014}, Logan, UT, USA, 2014.
  [Online]. Available:
  \url{http://digitalcommons.usu.edu/smallsat/2014/Propulsion/4/}
\BIBentrySTDinterwordspacing

\bibitem{Atchison2008}
J.~Atchison, ``{A passive microscale solar sail},'' in \emph{AIAA SPACE 2008
  Conference {\&} Exposition}, 2008, p. 7840.

\bibitem{Burkhardt2018}
Z.~Burkhardt, N.~Perakis, and C.~Welch, ``{Project Glowworm: Testing Laser Sail
  Propulsion in LEO},'' in \emph{International Astronautical Congress 2018},
  2018.

\bibitem{Streetman2016}
B.~Streetman, J.~Shoer, R.~Stoner, and M.~A. Peck, ``{Dual exploration
  architectures for breaking the decades-long cycles of planetary science},''
  in \emph{2016 IEEE Aerospace Conference}, 2016, pp. 1--15.

\bibitem{Streetman2015}
B.~Streetman, ``{Exploration Architecture with Quantum Inertial Gravimetry and
  In-situ ChipSat Sensors},'' NASA NIAC, Tech. Rep., 2015.

\bibitem{Colombo2013}
\BIBentryALTinterwordspacing
C.~Colombo, C.~L{\"{u}}cking, and C.~McInnes, ``{Orbit evolution, maintenance
  and disposal of SpaceChip swarms through electro-chromic control},''
  \emph{Acta Astronautica}, 2013. [Online]. Available:
  \url{http://www.sciencedirect.com/science/article/pii/S0094576512002354}
\BIBentrySTDinterwordspacing

\bibitem{Streetman2017}
B.~Streetman, J.~Shoer, and L.~Singh, ``{Limitations of scaling momentum
  control strategies to small spacecraft},'' in \emph{IEEE Aerospace
  Conference}, 2017, pp. 1--7.

\bibitem{Weis2016}
L.~Weis, ``{Chip-scale spacecraft swarms: Dynamics, control, and
  exploration},'' PhD Thesis, Cornell University, 2016.

\bibitem{Weis2016a}
L.~M. Weis and M.~A. Peck, ``{Dynamics of Chip-scale Spacecraft Swarms near
  Irregular Bodies},'' in \emph{54th AIAA Aerospace Sciences Meeting}, 2016, p.
  1468.

\bibitem{Lucking2012}
C.~L{\"{u}}cking, C.~Colombo, and C.~R. McInnes, ``{Electrochromic orbit
  control for smart-dust devices},'' \emph{Journal of Guidance, Control, and
  Dynamics}, vol.~35, no.~5, pp. 1548--1558, 2012.

\bibitem{Atchison2011}
J.~A. Atchison and M.~A. Peck, ``{Length scaling in spacecraft dynamics},''
  \emph{Journal of guidance, control, and dynamics}, vol.~34, no.~1, pp.
  231--246, 2011.

\bibitem{Atchison2010a}
J.~A. Atchison, Z.~R. Manchester, and M.~A. Peck, ``{Microscale atmospheric
  re-entry sensors},'' in \emph{International Planetary Probe Workshop}, 2010.

\bibitem{Vivenzio2019}
S.~Vivenzio, D.~Fries, and C.~Welch, ``{Feasibility and Preliminary Design of a
  ChipSat Planetary Entry Mission to Investigate the Atmosphere of Venus},'' in
  \emph{International Astronautical Congress 2019}, 2019.

\bibitem{Kaluthantrige2019}
A.~Kaluthantrige, A.~Hein, and C.~Welch, ``{Laser sail deployment mechanism for
  a Chipsat},'' International Space University, Tech. Rep., 2019.

\bibitem{Rocha2019}
S.~Rocha, M.~Langer, and C.~Welch, ``{Solar sail storage and deployment
  mechanism for ChipSats},'' International Space University, Tech. Rep., 2019.

\bibitem{ThumbSat2015}
ThumbSat, ``{ThumbSat},'' 2015.

\bibitem{Brashears2016}
T.~Brashears, P.~Lubin, N.~Rupert, E.~Stanton, A.~Mehta, P.~Knowles, and G.~B.
  Hughes, ``{Building the future of wafersat spacecraft for relativistic
  spacecraft},'' in \emph{Planetary Defense and Space Environment
  Applications}, vol. 9981.\hskip 1em plus 0.5em minus 0.4em\relax
  International Society for Optics and Photonics, 2016, p. 998104.

\bibitem{Thangavelautham2016}
J.~Thangavelautham, M.~Herreras-Martinez, A.~Warren, A.~Chandra, and
  E.~Asphaug, ``{The SunCube FemtoSat Platform: A Pathway to Low-Cost
  Interplanetary Exploration},'' in \emph{6th Interplanetary CubeSat Workshop},
  Oxford, UK, 2016.

\bibitem{Hein2016c}
A.~Hein, ``{Heritage Technologies in Space Programs - Assessment Methodology
  and Statistical Analysis},'' Ph.D. dissertation, PhD thesis, Technical
  University of Munich, 2016.

\bibitem{Savransky2000}
S.~D. Savransky, \emph{{Engineering of creativity: Introduction to TRIZ
  methodology of inventive problem solving}}.\hskip 1em plus 0.5em minus
  0.4em\relax CRC Press, 2000.

\bibitem{Montgomery2017}
E.~E. Montgomery and L.~Johnson, ``{Solar Sail Propulsion: A Roadmap from
  Today's Technology to Interstellar Sailships},'' NASA, Tech. Rep., 2017.

\bibitem{Larson1999a}
W.~J. Larson and J.~R. Wertz, \emph{{Space mission analysis and design}},
  3rd~ed.\hskip 1em plus 0.5em minus 0.4em\relax Microcosm, 1999.

\bibitem{Bruno2014}
A.~Bruno, E.~Maghsoudi, and M.~J. Martin, ``{Simulation of Heat Transfer in
  Wafer-Integrated Femtosatellites},'' \emph{Journal of Spacecraft and
  Rockets}, vol.~51, no.~2, pp. 627--631, 2014.

\bibitem{Wang2017}
Y.~Wang, W.~Ye, Y.~Hong, Y.~Cao, and Y.~Zheng, ``{Feasibility study: highly
  integrated chipset design for compact synthetic aperture radar payload on
  micro-satellite},'' \emph{Journal of Electromagnetic Waves and applications},
  vol.~31, no.~6, pp. 594--603, 2017.

\bibitem{Topol2006}
A.~Topol, D.~{La Tulipe}, L.~Shi, D.~Frank, K.~Bernstein, S.~Steen, A.~Kumar,
  G.~Singco, A.~Young, K.~Guarini, and M.~Ieong, ``{Three-dimensional
  integrated circuits},'' \emph{IBM Journal of Research and Development},
  vol.~50, no. 4.5, pp. 491--506, 2006.

\bibitem{Short2014}
K.~Short and D.~{Van Buren}, ``{Printable Spacecraft: Flexible Electronic
  Platforms for NASA Missions},'' NASA Innovative Advanced Concepts (NIAC)
  Phase 2], Tech. Rep., 2014.

\bibitem{Barnhart2006}
D.~J. Barnhart, T.~Vladimirova, and M.~N. Sweeting, ``{Satellite-on-a-Chip: A
  Feasibility Study},'' Tech. Rep., 2006.

\bibitem{Pister1999}
\BIBentryALTinterwordspacing
K.~Pister, J.~Kahn, and B.~Boser, ``{Smart dust: Wireless networks of
  millimeter-scale sensor nodes},'' \emph{Highlight Article in}, 1999.
  [Online]. Available:
  \url{https://scholar.google.fr/scholar?q=smart+dust{\&}btnG={\&}hl=de{\&}as{\_}sdt=0{\%}252C5{\#}9}
\BIBentrySTDinterwordspacing

\bibitem{Doherty2001}
\BIBentryALTinterwordspacing
L.~Doherty, B.~Warneke, B.~Boser, and K.~Pister, ``{Energy and performance
  considerations for smart dust},'' \emph{International Journal of Parallel and
  {\ldots}}, 2001. [Online]. Available:
  \url{https://scholar.google.fr/scholar?q=smart+dust{\&}btnG={\&}hl=de{\&}as{\_}sdt=0{\%}252C5{\#}4}
\BIBentrySTDinterwordspacing

\bibitem{Warneke2001a}
\BIBentryALTinterwordspacing
B.~Warneke, M.~Last, B.~Liebowitz, and K.~Pister, ``{Smart dust: Communicating
  with a cubic-millimeter computer},'' \emph{Computer}, vol.~34, no.~1, pp.
  44--51, 2001. [Online]. Available:
  \url{http://ieeexplore.ieee.org/xpls/abs{\_}all.jsp?arnumber=895117}
\BIBentrySTDinterwordspacing

\bibitem{Warneke2001}
\BIBentryALTinterwordspacing
B.~Warneke, B.~Atwood, and K.~Pister, ``{Smart dust mote forerunners},'' in
  \emph{14th IEEE International Conference on Micro Electro Mechanical
  Systems}, 2001, pp. 357--360. [Online]. Available:
  \url{http://ieeexplore.ieee.org/xpls/abs{\_}all.jsp?arnumber=906552}
\BIBentrySTDinterwordspacing

\bibitem{Niccolai2019}
L.~Niccolai, M.~Bassetto, A.~A. Quarta, and G.~Mengali, ``{A review of Smart
  Dust architecture, dynamics, and mission applications},'' \emph{Progress in
  Aerospace Sciences}, vol. 106, pp. Pages 1--14, 2019.

\bibitem{Waser2003}
R.~Waser, \emph{{Nanoelectronics and information technology}}.\hskip 1em plus
  0.5em minus 0.4em\relax Weilheim: Wiley-VCH, 2003.

\bibitem{Lu2010}
W.~Lu and C.~M. Lieber, ``{Nanoelectronics from the bottom up},'' in
  \emph{Nanoscience And Technology: A Collection of Reviews from Nature
  Journals}, 2010, pp. 137--146.

\bibitem{Akinwande2014}
D.~Akinwande, N.~Petrone, and J.~Hone, ``{Two-dimensional flexible
  nanoelectronics},'' \emph{Nature communications}, vol.~5, p. 5678, 2014.

\bibitem{Xiang2016}
D.~Xiang, X.~Wang, C.~Jia, T.~Lee, and X.~Guo, ``{Molecular-scale electronics:
  from concept to function},'' \emph{Chemical Reviews}, vol. 116, no.~7, pp.
  4318--4440, 2016.

\bibitem{Craighead2000}
\BIBentryALTinterwordspacing
H.~Craighead, ``{Nanoelectromechanical systems},'' \emph{Science}, vol. 290,
  no. 5496, pp. 1532--1535, 2000. [Online]. Available:
  \url{http://science.sciencemag.org/content/290/5496/1532.short}
\BIBentrySTDinterwordspacing

\bibitem{Meem2019}
M.~Meem, S.~Banerji, A.~Majumder, F.~G. Vasquez, B.~Sensale-Rodriguez, and
  R.~Menon, ``{Broadband lightweight flat lenses for longwave-infrared
  imaging},'' \emph{PNAS}, vol. 116, no.~43, pp. 21\,375--21\,378, 2019.

\bibitem{Eelkema2006}
R.~Eelkema, M.~Pollard, J.~Vicario, N.~Katsonis, B.~Ramon, C.~Bastiaansen,
  D.~Broer, and B.~Feringa, ``{Molecular machines: nanomotor rotates microscale
  objects},'' \emph{Nature}, vol. 440, no. 7081, p. 163, 2006.

\bibitem{Douglas2012}
S.~M. Douglas, I.~Bachelet, and G.~M. Church, ``{A logic-gated nanorobot for
  targeted transport of molecular payloads},'' \emph{Science}, vol. 335, no.
  6070, pp. 831--834, 2012.

\bibitem{Hein2017}
\BIBentryALTinterwordspacing
A.~Hein, K.~F. Long, D.~Fries, N.~Perakis, A.~Genovese, S.~Zeidler, M.~Langer,
  R.~Osborne, R.~Swinney, J.~Davies, B.~Cress, M.~Casson, A.~Mann, and
  R.~Armstrong, ``{The Andromeda Study: A Femto-Spacecraft Mission to Alpha
  Centauri},'' \emph{arXiv preprint}, vol. 1708.03556, 2017. [Online].
  Available: \url{http://arxiv.org/abs/1708.03556}
\BIBentrySTDinterwordspacing

\bibitem{Messerschmitt2012}
\BIBentryALTinterwordspacing
D.~Messerschmitt, ``{Interstellar communication: The case for spread
  spectrum},'' \emph{Acta Astronautica}, vol.~8, no.~1, pp. 227--238, 2012.
  [Online]. Available:
  \url{http://www.sciencedirect.com/science/article/pii/S0094576512002871}
\BIBentrySTDinterwordspacing

\bibitem{Matloff2005}
\BIBentryALTinterwordspacing
G.~Matloff and L.~Johnson, ``{Applications of the electrodynamic tether to
  interstellar travel},'' 2005. [Online]. Available:
  \url{http://ntrs.nasa.gov/search.jsp?R=20050215611}
\BIBentrySTDinterwordspacing

\end{thebibliography}

\end{document}